# Thermal Expansion and Volume Magnetostriction in Breathing Pyrochlore Magnets Li$A$Cr$_4$X$_8$ ($A$ = Ga, In, $X$ = O, S)

Tomoya Kanematsu[1], Masaki Mori[1], Yoshihiko Okamoto[1,*], Takeshi Yajima[2], and Koshi Takenaka[1]

[1]*Department of Applied Physics, Nagoya University, Nagoya 464-8603, Japan*
[2]*Institute for Solid State Physics, University of Tokyo, Kashiwa 277-8581, Japan*

We report thermal expansion and magnetostriction in breathing pyrochlore magnets Li$A$Cr$_4$X$_8$ ($A$ = Ga, In, $X$ = O, S) measured by a dilatometric method on sintered samples. All four of these compounds showed a large volume contraction associated with antiferromagnetic order with decreasing temperature. Above the Néel temperature, LiGaCr$_4$S$_8$ showed negative thermal expansion, LiInCr$_4$O$_8$ showed positive thermal expansion with concave-downward temperature dependence, and LiInCr$_4$S$_8$ showed positive forced volume magnetostriction. All these phenomena are likely caused by the complex structure-magnetism correlation within the breathing pyrochlore structure with $J$ and $J'$. These results suggested that breathing pyrochlore magnets are promising for the realization of various volumetric phenomena related to their magnetism not only in the magnetically-ordered phase but also in the paramagnetic phase.

Studies of correlation between the magnetic properties of solids and their volume have a long history spanning more than a century. The most traditional example is the magnetovolume effect in ferromagnetic metals. In the well-known invar alloy, the spontaneous volume magnetostriction associated with ferromagnetic order cancels the thermal expansion due to anharmonic lattice vibrations, resulting in nearly zero thermal expansion over a wide temperature range below 400 K.[1,2] With this characteristic, invar alloys are used as structural materials for precision machines. In antiferromagnetic metals, there are some materials that show a large volume change associated with antiferromagnetic order, such as antiperovskite-type Mn$_3$GaN and cubic Laves phase YMn$_2$.[3-5] In particular, in the former case, a large volume change is changed into a large negative thermal expansion by chemical substitution, which is of interest with regards to the control of thermal expansion.[6-8] Invar alloys also exhibit a strong lattice response to magnetic fields. In addition to the magnetostriction that generally appears in ferromagnets, they show a large volume change due to the application of magnetic fields (forced volume magnetostriction) in the ferromagnetic phase. The magnitude was reported to be $\Delta V/V$ = 220 ppm at room temperature when a magnetic field of 1600 Oe was applied, which is two orders of magnitude larger than that of Fe and Ni.[9]

Although there are less studies than on metals, some insulators are known to show correlated phenomena between magnetic properties and volume. Cr spinel compounds are a typical example. ZnCr$_2$Se$_4$ was reported to show not only a large volume contraction associated with antiferromagnetic order at $T_N$ = 21 K with decreasing temperature, but also negative thermal expansion and forced volume magnetostriction above $T_N$.[10] The importance of strong spin-lattice coupling and the competition between antiferro- magnetic and ferromagnetic interactions has been noted; however, the formation mechanism of these phenomena is less well understood than the magnetovolume effects in magnetic metals.

In order to elucidate the correlation between magnetic properties and volume in Cr spinel compounds, we focused on the $A$-site ordered Cr spinel compounds Li$A$Cr$_4$X$_8$ ($A$ = Ga, In, $X$ = O, S). In these compounds, each Cr atom is octahedrally coordinated by six O or S atoms and trivalent with three $3d$ electrons. They are localized at a Cr$^{3+}$ ion and occupy $xy$, $yz$, and $zx$ orbitals, respectively, resulting in an $S$ = 3/2 Heisenberg spin system without orbital degrees of freedom. Cr$^{3+}$ ions form a breathing pyrochlore structure, which consists of alternating small and large tetrahedra with magnetic interactions of $J$ and $J'$, respectively (see the inset of Fig. 1). Breathing pyrochlore magnets have recently attracted attention as promising systems for the realization of interesting magnetic properties caused by the coexistence of geometrical frustration and bond alternation.[11-15] LiGaCr$_4$O$_8$ and LiInCr$_4$O$_8$ are antiferromagnets with Weiss temperatures of $\theta_W$ = −659 and −332 K, respectively, and their $J$ and $J'$ are antiferromagnetic ($J$, $J'$ > 0).[11] The ratios of $J$ and $J'$ were estimated to be $J'/J$ ~ 0.6 and ~ 0.1 for $A$ = Ga and In, respectively, by using an empirical relationship between the strength of the magnetic interactions and the Cr–Cr bond lengths. LiInCr$_4$O$_8$ showed spin-gap behavior caused by spin-singlet formation in the small tetrahedra at low temperatures, but both compounds showed antiferromagnetic long-range order at $T_N \sim 15$ K.[16-19] A complex two-up two-down spin structure was suggested by neutron diffraction experiments on LiInCr$_4$O$_8$, while spin structure for LiGaCr$_4$O$_8$ is still unclear.[17] In addition, both compounds were reported to be insulating.[18] Various intriguing magnetic phenomena related to this magnetically-ordered phase have been observed, such as the tricritical point in the magnetic phase diagram,[16] a transition to the possible spin-nematic phase,[20] and a half-magnetization plateau under strong magnetic fields.[21,22]

The sulfide Li$A$Cr$_4$S$_8$ is isostructural to the oxide, but unlike the oxide, antiferromagnetic and ferromagnetic interactions coexist and compete each other, with θ$_W$ values of −20 and 30 K for $A$ = Ga and In, respectively. These small |θ$_W$| values suggest that both compounds have antiferromagnetic $J$ and ferromagnetic $J'$.[23] On the other hand, a theoretical study using first-principles calculations reported that both $J$ and $J'$ are ferromagnetic and the further neighbor interactions are antiferromagnetic in Li$A$Cr$_4$S$_8$.[24] Although the average magnetic interaction in LiInCr$_4$S$_8$ is ferromagnetic, a magnetic phase transition accompanied by a discontinuous decrease of magnetic susceptibility with decreasing temperature appears at 24 K, suggesting that an antiferromagnetic order accompanied by a structural change occurs at this temperature.[23,25] Spin structure in the magnetically ordered phase of LiInCr$_4$S$_8$ is still unclear. In LiGaCr$_4$S$_8$, magnetic susceptibility shows a peak at ~ 10 K, which might be due to antiferromagnetic order.[23] The electrical resistivity of LiGaCr$_4$S$_8$ was reported to be semiconducting.[26]

Thus, Li$A$Cr$_4$X$_8$ has been studied mainly from the viewpoint of magnetism, with few studies reporting volumetric properties, such as thermal expansion. The structural parameters of Li$A$Cr$_4$X$_8$ at low temperatures determined by crystallographic methods were reported in several previous studies.[17,19,26] In particular, powder X-ray and neutron diffraction data of LiGaCr$_4$S$_8$ indicated that negative thermal expansion occurs in the paramagnetic phase between 10 and 110 K.[26] Moreover, no studies have focused on the volumetric properties of Li$A$Cr$_4$X$_8$ using dilatometric methods. In this study, we report the correlation between the volumetric and magnetic properties of Li$A$Cr$_4$X$_8$ revealed by comprehensive thermal expansion and magnetostriction measurements using sintered samples. We found various intriguing magnetovolume phenomena in Li$A$Cr$_4$X$_8$, such as anomalous thermal expansion probably due to the magnetism in the paramagnetic phase and forced volume magnetostriction in antiferromagnets.

Polycrystalline samples of Li$A$Cr$_4$X$_8$ ($A$ = Ga, In, $X$ = O, S) were prepared by a solid-state reaction method described in Refs. 11 and 23. The obtained polycrystalline samples were sintered using a spark plasma sintering furnace (SPS Syntex). Linear thermal expansion and linear strain in magnetic fields of the sintered samples were measured using a strain gage (KFLB, 120 W, Kyowa Electronic Instruments Co.) with a Cu reference.[8] Linear strain in magnetic fields, heat capacity, and some of linear thermal-expansion measurements were performed using a Physical Property Measurement System (Quantum Design). Powder X-ray diffraction patterns of a LiGaCr$_4$S$_8$ sample at various temperatures were measured by using a SmartLab diffractometer (Rigaku) with Cu Kα1 radiation, monochromated by a Ge(111)-Johansson-type monochromator.

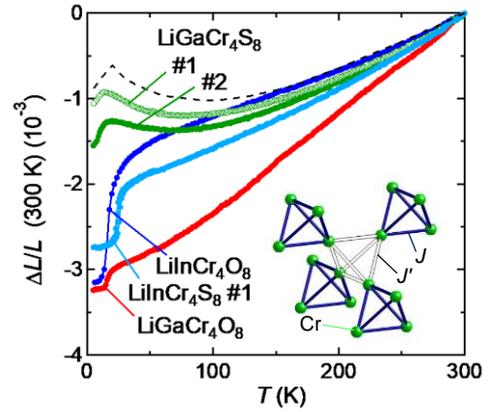

Fig. 1. Temperature dependence of linear thermal expansion of Li$A$Cr$_4$X$_8$ ($A$ = Ga, In, $X$ = O, S) sintered samples normalized to the 300 K data. The data for samples #1 and #2 are shown for LiGaCr$_4$S$_8$, and the data for sample #1 are shown for LiInCr$_4$S$_8$. The broken line indicates the linear thermal expansion of the sample #1 of LiGaCr$_4$S$_8$ calculated by using the lattice parameters determined by powder X-ray diffraction. The inset shows breathing pyrochlore structures made by Cr atoms in Li$A$Cr$_4$X$_8$. The tetrahedra with filled and open lines indicate the small and large tetrahedra with $J$ and $J'$, respectively.

Figure 1 shows the temperature dependence of linear thermal expansion $\Delta L/L$ of the sintered samples of Li$A$Cr$_4$X$_8$ ($A$ = Ga, In, $X$ = O, S). All samples showed anomalies at low temperatures. LiGaCr$_4$O$_8$ with the strongest antiferromagnetic interaction among the four compounds showed positive thermal expansion over the entire measured temperature range between 300 and 5 K. With decreasing temperature, a rapid contraction of $\Delta L/L$ = 200 ppm occurs between 19 and 14 K across $T_N$, followed by an almost constant value below 14 K. LiInCr$_4$O$_8$ with a smaller $J'/J$ compared to LiGaCr$_4$O$_8$ also exhibited positive thermal expansion between 300 and 5 K, but it shows concave-downward temperature dependence below ~70 K, corresponding to the temperature region where the spin-gap behavior appeared in magnetic susceptibility data.[11] At lower temperatures, the $\Delta L/L$ of LiInCr$_4$O$_8$ shows a sharp drop at around $T_N$ = 15 K with decreasing temperature, and becomes almost constant below $T_N$.

The linear thermal expansion $\Delta L/L$ of LiInCr$_4$S$_8$ showed similar behavior to that of the most antiferromagnetic LiGaCr$_4$O$_8$, although LiInCr$_4$S$_8$ has a positive θ$_W$ and is the most ferromagnetic in the four compounds. Positive thermal expansion was observed between 300 and 5 K, and a strong contraction approaching $\Delta L/L$ = 700 ppm occurred between 30 and 20 K across $T_N$, with decreasing temperature. In contrast, LiGaCr$_4$S$_8$ showed negative thermal expansion between 80 and 20 K. As shown in Fig. 1, the total volume change caused by the negative thermal expansion is $\Delta V/V$ = 800 and 330 ppm for sample #1 and #2, respectively. At lower temperatures, $\Delta L/L$ shows a peak at ~10 K, where magnetic susceptibility shows a peak, and then shows positive thermal expansion. As seen in Fig. 1, the values of $\Delta L/L$ and the temperature region, where negative thermal expansion occurred, are almost consistent with those estimated using the lattice constants determined by powder X-ray diffraction measurements. This result indicates that the microstructural effect discussed for some negative-thermal-expansion materials, such as β-eucryptite and Ca$_2$RuO$_4$, does not work and the observed negative thermal expansion is a bulk property.[27-29]

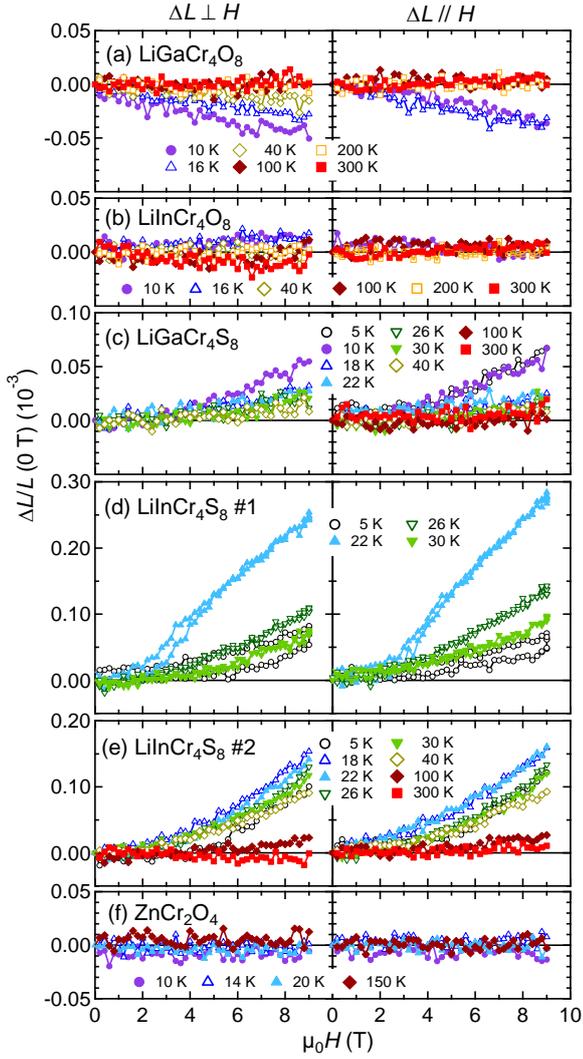

Fig. 2. Magnetic field dependences of linear strain of the sintered samples of Li$A$Cr$_4$X$_8$ ($A$ = Ga, In, $X$ = O, S) and ZnCr$_2$O$_4$. All the data are normalized to the zero-field data. The left and right panels show transverse and longitudinal magnetostrictions, where linear strains were measured perpendicular to and along magnetic fields, respectively. For LiInCr$_4$S$_8$, the data of sample #1 and #2 are shown. The data for LiGaCr$_4$S$_8$ correspond to sample #2 in Fig. 1.

Figure 2 shows magnetic field dependences of the linear strains $\Delta L/L$ of Li$A$Cr$_4$X$_8$ sintered samples ($A$ = Ga, In, $X$ = O, S) measured at various temperatures. The data for transverse magnetostriction, where the linear strain was measured perpendicular to the magnetic field, and those for longitudinal magnetostriction, where the linear strain was measured parallel to the magnetic field, are shown. In Fig. 2(f), those of a sintered sample of ZnCr$_2$O$_4$ with a uniform pyrochlore structure are shown as a reference. The transverse and longitudinal magnetostrictions show similar behaviors for all the measured samples, indicating that there is no significant anisotropy in the linear strain data. As shown in Fig. 2(a), the data of LiGaCr$_4$O$_8$ above 100 K do not show significant magnetostriction up to a magnetic field of 9 T. At lower temperatures, however, linear strain $\Delta L/L$ decreases with increasing magnetic fields. The magnitude of contraction at 10 K is $\Delta L/L$ ~ 40 ppm between zero magnetic field and 9 T, regardless of the direction of the magnetic field, corresponding to a volume contraction of $\Delta V/V$ ~ 120 ppm. This negative forced volume magnetostriction also appeared at 16 and 40 K above $T_N$. In contrast, as shown in Fig. 2(b), there is no anomaly in all the data of LiInCr$_4$O$_8$, which is similar to the behavior in ZnCr$_2$O$_4$ shown in Fig. 2(f).

In contrast to the oxides, the sulfide samples show positive forced volume magnetostriction. As seen in Fig. 2(d), sample #1 of LiInCr$_4$S$_8$ shows a strong increase of linear strain $\Delta L/L$ above ~3 T at 22 K just below $T_N$. $\Delta L/L$ between the zero magnetic field and 9 T is ~250 and ~280 ppm for transverse and longitudinal magnetostrictions, respectively, yielding $\Delta V/V$ = 780 ppm. Although not as pronounced as the 22 K data, significant positive linear strain was observed at 5, 26, and 30 K. In contrast, the linear strain $\Delta L/L$ of sample #2 shown in Fig. 2(e) is considerably smaller than that of sample #1. The LiGaCr$_4$S$_8$ sample shows positive volume magnetostriction of $\Delta L/L$ = 70 and 50 ppm for transverse and longitudinal magnetostrictions between 0 and 9 T at 10 K, respectively.

Here, we discuss the thermal expansion and volume magnetostriction in the four breathing pyrochlore magnets. First, as seen in Fig. 1, all four compounds showed rapid volume contraction at around $T_N$ with decreasing temperature. Similar volume contractions have been reported in many Cr spinels with a uniform pyrochlore structure,[10,30-32] indicating that the volume contraction at around $T_N$ with decreasing temperature appears regardless of the breathing. In the Cr spinel compounds, spin frustration is relieved by distorting the pyrochlore structure at low temperatures, resulting in antiferromagnetic order and structural distortion occurring simultaneously. This phase transition is called the spin-Jahn-

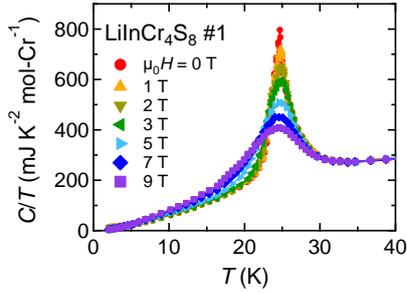

Fig. 3. Temperature dependence of heat capacity divided by temperature of the sintered sample of $LiInCr_4S_8$ under various magnetic fields. The sample corresponds to sample #1 in Fig. 2.

Teller transition.[33] Since the pyrochlore structure is a cage-like structure with large pores, the cages are always slightly crushed when the structural distortion occurs, giving rise to the contraction of the unit cell volume.

The large and positive forced volume magnetostriction above 3 T in sample #1 of $LiInCr_4S_8$ at 22 K just below $T_N$, shown in Fig. 2(d), is related to the antiferromagnetic order accompanied by structural distortion discussed above. The antiferromagnetic order is destroyed by applying a magnetic field, which relieves the structural distortion of the pyrochlore structure and expands the unit cell volume. This is supported by the heat capacity data shown in Fig. 3, where the peak in heat capacity divided by temperature $C/T$ becomes broader and shifts to lower temperature with increasing magnetic field. Similar volume expansion was reported in $ZnCr_2Se_4$.[10] $LiGaCr_4S_8$ and sample #2 of $LiInCr_4S_8$ show smaller and more continuous volume expansion than that in sample #1 of $LiInCr_4S_8$, probably due to lower crystallinity and broader magnetic transition in the former two samples. $LiGaCr_4O_8$ and $LiInCr_4O_8$ do not show significant volume expansion by applying a magnetic field, as shown in Figs. 2(a) and (b), although they show rapid volume contraction at around $T_N$ with decreasing temperature, similar to sulfides. The robustness of the $T_N$ to the magnetic field due to the strong antiferromagnetic interactions is likely responsible for the absence of volume expansion in magnetic fields in oxides.[11]

$LiACr_4X_8$ is interesting in that it also exhibited various volumetric phenomena above $T_N$. First, we discuss the negative thermal expansion in $LiGaCr_4S_8$. This negative thermal expansion is likely of magnetic origin, because it completely disappeared below the magnetic transition temperature of 10 K.[23] Recently, the negative thermal expansion of a series of antiperovskite-type Mn nitrides was explained by Mochizuki et al.[34,35] They reported that the negative thermal expansion appears when antiferromagnetic and ferromagnetic interactions $J_{AF}$ and $J_F$ have different bond length dependences and compete each other ($0.7J_{AF} < |J_F| < J_{AF}$ for Mn nitrides). It is likely that only $LiGaCr_4S_8$ among the four compounds satisfies this condition, and thus only this compound exhibits a negative thermal expansion, whereas the other three compounds do not.[23] However, the negative thermal expansion in $LiGaCr_4S_8$ appeared in the paramagnetic phase above $T_N$, which is different from that in the Mn nitrides, where the negative thermal expansion is associated with the antiferromagnetic order. $ZnCr_2Se_4$ and $CdCr_2O_4$ were reported to show negative thermal expansion in a paramagnetic phase above $T_N$.[10,30] Such a negative thermal expansion in a paramagnetic phase is expected to increase its upper temperature limit, which might open an avenue for finding novel negative thermal expansion materials for practical use.

In contrast to $LiGaCr_4S_8$, the volume of the $LiInCr_4O_8$ sample strongly decreases with decreasing temperature above $T_N$. In $LiInCr_4O_8$, both $J$ and $J'$ are antiferromagnetic, but because of the large difference between $J$ and $J'$, $LiInCr_4O_8$ showed spin-gap-like behavior above $T_N$. Since the temperature range where spin-gap-like behavior appears in magnetic susceptibility and a strong volume contraction occurs are almost identical, these two phenomena are most likely related. Perhaps, the smaller tetrahedron, in which the spin singlet is formed, becomes smaller to make the antiferromagnetic interaction stronger with decreasing temperature.

Finally, we discuss the forced volume magnetostriction appearing above $T_N$. As seen in Figs. 2(d) and (e), $LiInCr_4S_8$ showed a positive forced volume magnetostriction above $T_N$. Similar positive magnetostriction was also observed in $ZnCr_2Se_4$.[10] Both compounds are similar in that the average magnetic interaction is ferromagnetic, as indicated by positive $\theta_W$. In the Cr spinel compounds, the magnetic interaction between the nearest neighbor Cr spins becomes more ferromagnetic with increasing Cr–Cr bond length. With increasing magnetic field, the Cr–Cr bond length in the large tetrahedron with ferromagnetic $J'$ in $LiInCr_4S_8$ is increased to gain energy. Thus, $LiACr_4X_8$ is intriguing in terms of not only a large volume change associated with magnetic ordering, but also various magnetovolume phenomena in the paramagnetic phase above $T_N$, due to the competition of magnetic interactions and strong coupling between structural parameters and magnetic properties.

In summary, we report thermal expansion and magnetostriction in four breathing pyrochlore magnets $LiACr_4X_8$ ($A$ = Ga, In, $X$ = O, S) measured by a dilatometric method on sintered samples. All four compounds showed a large volume contraction at around $T_N$ with decreasing temperature, which is due to structural distortion associated with antiferromagnetic long-range order. In sulfides, this effect also appeared as a positive forced volume magneto- striction. Moreover, these compounds showed various volumetric phenomena related to their magnetism in their paramagnetic phases. The coexistence of ferromagnetic and antiferromagnetic interactions on the breathing pyrochlore structures and the strong structure-magnetism correlation are likely to play important roles. The above results suggest that the breathing pyrochlore magnets are promising to realize various magnetovolume phenomena not only in the magnetically ordered phase but also in the paramagnetic phase.

## ACKNOWLEDGMENTS

This work was partly carried out under the Visiting Researcher Program of the Institute for Solid State Physics, the University of Tokyo and supported by JSPS KAKENHI (Grant Number: 19H05823, 20H02603).

*E-mail: yokamoto@nuap.nagoya-u.ac.jp